\documentclass[aps, superscriptaddress, txfonts, preprint]{revtex4-1}
\usepackage{textcomp}
\usepackage{amsmath}
\usepackage{graphicx}
\begin{document}

\title{Magnetic-field-induced suppression of spin Peltier effect in Pt/${\rm Y_{3}Fe_{5}O_{12}}$
system at room temperature}

\author{Ryuichi Itoh}

\affiliation{Institute for Materials Research, Tohoku University, Sendai 980-8577,
Japan}

\author{Ryo Iguchi}
\email{IGUCHI.Ryo@nims.go.jp}

\affiliation{Institute for Materials Research, Tohoku University, Sendai 980-8577,
Japan}

\affiliation{National Institute for Materials Science, Tsukuba 305-0047, Japan}

\author{Shunsuke Daimon}

\affiliation{Institute for Materials Research, Tohoku University, Sendai 980-8577,
Japan}

\affiliation{WPI Advanced Institute for Materials Research, Tohoku University,
Sendai 980-8577, Japan}

\author{Koichi Oyanagi}

\affiliation{Institute for Materials Research, Tohoku University, Sendai 980-8577,
Japan}

\author{Ken-ichi Uchida}

\affiliation{National Institute for Materials Science, Tsukuba 305-0047, Japan}

\affiliation{PRESTO, Japan Science and Technology Agency, Saitama 332-0012, Japan}

\affiliation{Center for Spintronics Research Network, Tohoku University, Sendai
980-8577, Japan}

\author{Eiji Saitoh}

\affiliation{Institute for Materials Research, Tohoku University, Sendai 980-8577,
Japan}

\affiliation{WPI Advanced Institute for Materials Research, Tohoku University,
Sendai 980-8577, Japan}

\affiliation{Center for Spintronics Research Network, Tohoku University, Sendai
980-8577, Japan}

\affiliation{Advanced Science Research Center, Japan Atomic Energy Agency, Tokai
319-1195, Japan}
\begin{abstract}
We report the observation of magnetic-field-induced suppression of
the spin Peltier effect (SPE) in a junction of a paramagnetic metal
Pt and a ferrimagnetic insulator ${\rm Y_{3}Fe_{5}O_{12}}$ (YIG)
at room temperature. For driving the SPE, spin currents are generated
via the spin Hall effect from applied charge currents in the Pt layer,
and injected into the adjacent thick YIG film. The resultant temperature
modulation is detected by a commonly-used thermocouple attached to
the Pt/YIG junction. The output of the thermocouple shows sign reversal
when the magnetization is reversed and linearly increases with the
applied current, demonstrating the detection of the SPE signal. We
found that the SPE signal decreases with the magnetic field. The observed
suppression rate was found to be comparable to that of the spin Seebeck
effect (SSE), suggesting the dominant and similar contribution of
the low-energy magnons in the SPE as in the SSE.
\end{abstract}
\maketitle

\section{Introduction}

Thermoelectric conversion is one of the promising technologies for
smart energy utilization \cite{Zhang201592}. Owing to the progress
of spintronics in this decade, the spin-based thermoelectric conversion
is now added to the scope of the thermoelectric technology \cite{Bauer:2012fq,boona2014spin,Uchida:2016jo,Flipse:2014cl,Flipse:2012kn,Slachter:2010hj}.
In particular, the thermoelectric generation mediated by flow of spins,
or spin current, has attracted much attention because of the advantageous
scalability, simple fabrication processes, and flexible design of
the figure of merit \cite{Uchida:2008cc,UchidaXiaoAdachiEtAl2010,Kirihara:2012jq,Ramos:2015kh,Uchida:2016jo,Adachi:2013jy}.
This is realized by combining the spin Seebeck effect (SSE) \cite{Uchida:2010jb}
and spin-to-charge conversion effects \cite{Saitoh:2006kk,Sinova:2015ic,Hoffmann:-1el},
where a spin current is generated by an applied thermal gradient and
is converted into electricity owing to spin\textendash orbit coupling.

The SSE has a reciprocal effect called the spin Peltier effect (SPE),
discovered by Flipse \textit{et al}. in 2014 in a Pt/yttrium iron
garnet (${\rm Y_{3}Fe_{5}O_{12}}$: YIG) junction \cite{Flipse:2014cl,Daimon:2016fja}.
In the SPE, a spin current across a normal conductor (N)/ferromagnet
(F) junction induces a heat current, which can change the temperature
distribution around the junction system.

To reveal the mechanism of the SPE, systematic experiments have been
conducted \cite{Uchida:2017kb,Daimon:2017jx,Daimon:2016fja}. Since
the SPE is driven by magnetic fluctuations (magnons) in the F layer,
detailed study on the magnetic-field and temperature dependence is
indispensable for clarifying the microscopic relation between the
SPE and magnon excitation and the reciprocity between the SPE and
SSE \cite{Uchida:2014jq,Rezende:2014cr,Kikkawa:2015bn,Jin:2015ik,Barker:2016hy,Guo:2016go,Basso:2016gc,Anonymous:2017jm,Anonymous:2017ju,ohnumxiv}.
A high magnetic field is expected to affect the magnitude of the SPE
signal via the modulation of spectral properties of magnons. In fact,
the SSE thermopower in a Pt/YIG system was shown to be suppressed
by high magnetic fields even at room temperature against the conventional
theoretical expectation based on the equal contribution over the magnon
spectrum \cite{Kikkawa:2015bn}. This anomalously-large suppression
highlights the dominant contribution of sub-thermal magnons, which
possess lower energy and longer propagation length than thermal magnons
\cite{Anonymous:2017de,Jin:2015ik,Guo:2016go,Anonymous:2017ju}. Thus,
the experimental examination of the field dependence of the SPE is
an important task for understanding the SPE. Although the SPE has
recently been measured in various systems by using the lock-in thermography
(LIT) \cite{Daimon:2016fja}, it is difficult to be used at high fields
and/or low temperatures. For investigating the high-magnetic-field
response of the SPE, an alternative method is required. 

In this paper, we investigate the magnetic field dependence of the
SPE up to 9 T at 300 K by using a commonly-used thermocouple (TC)
wire. As revealed by the LIT experiments \cite{Daimon:2016fja,Daimon:2017jx},
the temperature modulation induced by the SPE is localized in the
vicinity of N/F interfaces. This is the reason why the magnitude of
the SPE signals is very small in the first experiment by Flipse \textit{et
al}. \cite{Flipse:2014cl}, where a thermopile sensor is put on the
bare YIG surface, not on the Pt/YIG junction. Here, we show that the
SPE can be detected with better sensitivity simply by attaching a
common TC wire on a N/F junction. This simple SPE detection method
enables systematic measurements of the magnetic field dependence of
the SPE, since it is easily integrated to conventional measurement
systems. In the following, we describe the details of the electric
detection of the SPE signal using a TC, the results of the magnetic
field dependence of the SPE signal in a high-magnetic-field range,
and its comparison to that of the SSE thermopower.

\section{Experimental}

\begin{figure}
\includegraphics{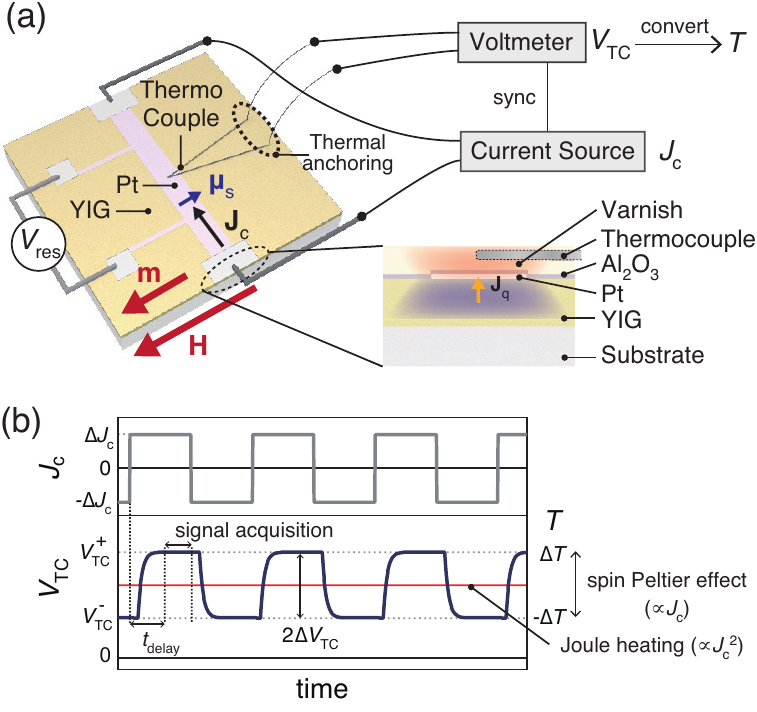}

\caption{(a) Schematic of the Pt/YIG sample and measurement system with a thermocouple
wire. $\mathbf{J}_{{\rm c}}$, $\boldsymbol{\mu}_{{\rm s}}$, $\mathbf{m}$,
$\mathbf{H}$, and $\mathbf{J}_{{\rm q}}$ denote the applied current,
the spin accumulation, the unit vector of the equilibrium magnetization,
the magnetic field, and the heat current concomitant with the spin-current
injection. In an isolated system, $\mathbf{J}_{{\rm q}}$ induces
a temperature gradient by accumulating heat, which can be detected
by the thermocouple. The resistance of the Pt strip is obtained by
measuring $V_{{\rm res}}$. The measurements were carried out by using
the physical property measurement system, Quantum Design. (b) Expected
responses due to the SPE and Joule heating when $J_{{\rm c}}$ is
periodically changed from $\Delta J_{{\rm c}}$ to $-\Delta J_{{\rm c}}$
with the zero offset current $J_{{\rm c}}^{0}=0$. Corresponding voltage
signals $V_{{\rm TC}}^{+}$ and $V_{{\rm TC}}^{-}$ are obtained after
the time delay $t_{{\rm delay}}$. The SPE signal is extracted from
the difference $\Delta V_{{\rm TC}}=\left(V_{{\rm TC}}^{+}-V_{{\rm TC}}^{-}\right)/2$
. \label{fig:Schematic}}
\end{figure}

The spin current for driving the SPE is generated via the spin Hall
effect (SHE) from a charge current applied to N  \cite{Sinova:2015ic,Hoffmann:-1el}.
The SHE-induced spin current then forms spin accumulation at the N/F
interface, whose spin vector representation is given by
\begin{equation}
\boldsymbol{\mu}_{{\rm s}}\propto\theta_{{\rm SHE}}\mathbf{j}_{{\rm c}}\times\mathbf{n},\label{eq:SHE}
\end{equation}
where $\theta_{{\rm SHE}}$ is the spin Hall angle of N, $\mathbf{j}_{{\rm c}}$
the charge-current-density vector, and $\mathbf{n}$ the unit vector
normal to the interface directing from F to N. $\boldsymbol{\mu}_{{\rm s}}$
at the interface exerts spin-transfer torque to magnons in F via the
interfacial exchange coupling at finite temperatures, when $\boldsymbol{\mu}_{{\rm s}}$
is parallel or anti-parallel to the equilibrium magnetization ($\mathbf{m}$)
\cite{Tserkovnyak:2005fr,Zhang:2012hh}. The torque increases or decreases
the number of the magnons depending on the polarization of the torque
($\boldsymbol{\mu}_{{\rm s}}\parallel\mathbf{m}$ or $\boldsymbol{\mu}_{{\rm s}}\parallel-\mathbf{m}$),
and eventually changes the system temperature by energy transfer,
concomitant with the spin-current injection \cite{Flipse:2014cl,Daimon:2016fja,ohnumxiv}.
The energy transfer induces observable temperature modulation in isolated
systems, which satisfies the following relation 
\begin{equation}
\Delta T_{{\rm SPE}}\propto\boldsymbol{\mu}_{{\rm s}}\cdot\mathbf{m}\propto\left(\mathbf{j}_{{\rm c}}\times\mathbf{n}\right)\cdot\mathbf{m}.\label{eq:SPE}
\end{equation}

A schematic of the sample system and measurement geometry is shown
in Fig. 1(a). The sample system is a Pt strip on a single-crystal
YIG. The YIG layer is 112-${\rm \mu m}$-thick and grown by a liquid
phase epitaxy method on a 500-${\rm \mu m}$-thick ${\rm Gd_{3}Ga_{5}O_{12}}$
substrate with the lateral dimension 10 \texttimes{} 10 ${\rm mm}{}^{2}$,
where small amount of Bi is substituted for the Y-site of the YIG
to compensate the lattice mismatching to the substrate. The Pt strip,
connected to four electrodes, is 5-nm-thick, 0.5-mm-wide, fabricated
by a sputtering method, and patterned with a metal mask. Then, the
whole surface of the sample except the electrodes is covered by a
highly-resistive ${\rm Al_{2}O_{3}}$ film with a thickness of $\sim100$
nm by means of an atomic layer deposition method. We attached a TC
wire to the top of the Pt/YIG junction, where the wire is electrically
insulated from but thermally connected to the Pt layer owing to the
${\rm Al_{2}O_{3}}$ layer. We used a type-E TC with a diameter of
0.013 mm (Omega engineering CHCO-0005), and fixed its junction part
on the middle of the Pt strip using varnish. The rest of the TC wires
were fixed on the sample surface for thermal-anchoring and avoiding
thermal leakage from the top of the Pt/YIG junction {[}see the cross
sectional view in Fig. \ref{fig:Schematic}(a){]}. The expected thickness
of the varnish between the TC and the sample surface is in the order
of $10\ \mu{\rm m}$ \footnote{This is estimated from the thickness of the varnish layer sandwiched
between glass substrates pressed with the same pressure applied to
the sample; we pressed the sample and the TC wire with an additional
glass cover.}. Note that the thickness of the varnish layer does not affect the
magnitude of the signal while it affects the temporal response of
the TC \footnote{As the radiation to the outer environment at the surface is negligibly
small, the vertical heat current in the varnish layer is zero at the
steady-state condition. The effect of the lateral heat currents, expected
at the edges of the Pt strip, is also small as the total thickness
from the top of the Pt strip to the surface ($\sim30\ {\rm \mu m}$)
is smaller than the width ($500\ {\rm \mu m}$). }. The ends of the Pt strip are connected to a current source and the
other two electrodes are used for measuring resistance based on the
four-terminal method. The magnetic field $\mathbf{H}$ is applied
in the film plane and perpendicular to the strip, thus is along $\boldsymbol{\mu}_{{\rm s}}$,
satisfying the symmetry of the SPE {[}Eq. \eqref{eq:SPE}{]}. The
TC is connected to electrodes on a heat bath, which acts as a thermal
anchor, and further connected to a voltmeter via conductive wires.
The measurements were carried out at 300 K and $\sim10^{-3}$ Pa.

For the electric detection of the SPE, we measured the amplitude difference
($\Delta V_{{\rm TC}}$) of the TC voltage $V_{{\rm TC}}$ in response
to a change ($\Delta J_{{\rm c}}$) in the current $J_{{\rm c}}$
(so-called Delta-mode of the nanovoltmeter Keithley 2182A). After
the current is set to $J_{{\rm c}}=J_{{\rm c}}^{0}\pm\Delta J_{{\rm c}}$,
time delay ($t_{{\rm delay}}$) is inserted before measuring the corresponding
voltages $V_{{\rm TC}}^{\pm}$. Then, $\Delta V_{{\rm TC}}$ is obtained
as $\Delta V_{{\rm TC}}=\left(V_{{\rm TC}}^{+}-V_{{\rm TC}}^{-}\right)/2$.
The time delay $t_{{\rm delay}}$ is necessary because the temperature
modulation occurred at the Pt/YIG junction takes certain time to reach
and stabilize the TC. The appropriate value of $t_{{\rm delay}}$
can be determined from the delay-dependence of the SPE signal, which
will be shown in Sec. III. For the SPE measurements, we set no offset
current ($J_{{\rm c}}^{0}=0$) so that $\Delta V_{{\rm TC}}$ is free
from Joule heating ($\propto J_{{\rm c}}^{2}$), and the SPE $\left(\propto J_{{\rm c}}\right)$
is expected to dominate the $\Delta V_{{\rm TC}}$ signal {[}see Fig.
\ref{fig:Schematic}(a){]}. By using the measured values of $\Delta V_{{\rm TC}}$,
the temperature modulation ($\Delta T$) is estimated via the relation
$\Delta T=S_{{\rm TC}}\Delta V_{{\rm TC}}$, where $S_{{\rm TC}}$
is the Seebeck coefficient of the TC. For the low-field measurements
($\mu_{0}H<0.1\ {\rm T}$), the reference value of $S_{{\rm TC}}=61\ {\rm \mu V/K}$
at 300 K is used, while, for the high-field measurements, the field
dependence of $S_{{\rm TC}}$, determined by the method shown in Appendix
\ref{sec:Calibration-of-Thermo}, is used.

\section{Results and Discussion}

\begin{figure}
\includegraphics{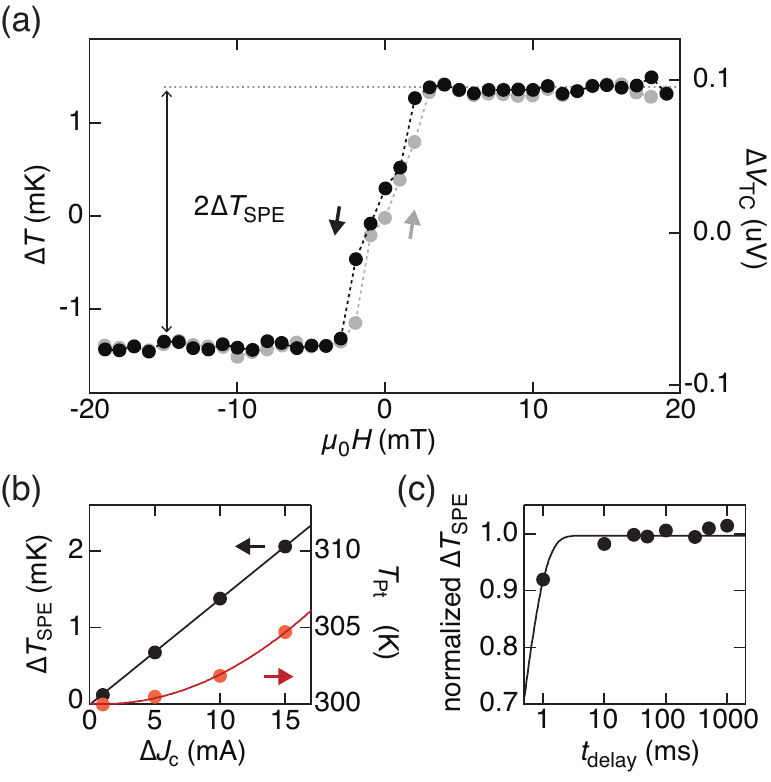}

\caption{(a) Field magnitude $H$ dependence of $\Delta V_{{\rm TC}}$ (the
right axis) and $\Delta T$ without offset (the left axis), measured
at $\Delta J_{{\rm c}}=10\ {\rm mA}$ and $t_{{\rm delay}}=50\ {\rm ms}$.
(b) Current $J_{{\rm c}}$ dependence of $\Delta T_{{\rm SPE}}$ at
$\mu_{0}\left|H\right|=0.1\ {\rm T}$, where $\mu_{0}$ represents
the permeability of vacuum. (c) Delay-time $t_{{\rm delay}}$ dependence
of $\Delta T_{{\rm SPE}}$. Plotted data are estimated from the results
measured at $\Delta J_{{\rm c}}=10\ {\rm mA}$ and $5\ {\rm mT}<\mu_{{\rm 0}}\left|H\right|<20\ {\rm mT}$
and normalized based on the fitting with $B\left[1-{\rm exp}\left(-t_{{\rm delay}}/\tau\right)\right]$,
where $B$ denotes the proportional constant and $\tau=0.4\ {\rm ms}$
denotes the characteristic time scale. The acquisition time of 20
ms is used for measurements.\label{fig:LF-SPE}}
\end{figure}

First, we demonstrated the electric detection of the SPE at low fields.
Figure \ref{fig:LF-SPE}(a) shows $\Delta V_{{\rm TC}}$ as a function
of the field magnitude $H$ at $\Delta J_{{\rm c}}=10\ {\rm mA}$
and $t_{{\rm delay}}=50\ {\rm ms}$. $\Delta V_{{\rm TC}}$ clearly
changes its sign when the field direction is reversed and the appearance
of the hysteresis demonstrates that it reflects the magnetization
curve of the YIG, showing the symmetry expected from Eq. \eqref{eq:SPE}
\cite{Daimon:2016fja,Flipse:2014cl}. The small offset of $\Delta V_{{\rm TC}}$
may be attributed to the temperature modulation by the Peltier effect
appearing around the current electrodes, Joule heating due to small
uncanceled current offsets, and possible electrical leakage of the
applied current from the sample to the TC. Since the Peltier and resistance
effects are of even functions of the magnetization cosines though
the SPE is of an odd function, the SPE-induced temperature modulation
$\Delta T_{{\rm SPE}}$ can be extracted by subtracting the symmetric
response to the magnetization: $\Delta T_{{\rm SPE}}=\left[\Delta T\left(+H\right)-\Delta T\left(-H\right)\right]/2$
\cite{Flipse:2014cl}. Figure \ref{fig:LF-SPE}(b) shows the $\Delta J_{{\rm c}}$
dependence of $\Delta T_{{\rm SPE}}$ and the temperature ($T_{{\rm Pt}}$)
of the Pt strip, estimated from the resistance of the strip. While
$T_{{\rm Pt}}$ increases parabolically with the magnitude of $\Delta J_{{\rm c}}$
for Joule heating, $\Delta T{\rm _{SPE}}$ increases linearly as is
expected from the characteristic of the SPE. This distinct dependencies
show negligibly small contribution to $\Delta T_{{\rm SPE}}$ from
the Joule heating in this study \footnote{The SPE signal at 305 and 310 K (nominal) was observed to show the
same magnitude as that at 300 K. Thus the temperature increase due
to the Joule heating does not affect the measured SPE value.}. The magnitude of the SPE signal is estimated to be $\Delta T_{{\rm SPE}}/\Delta j_{{\rm c}}=3.4\times10^{-13}\ {\rm Km^{2}/A}$,
where $\Delta j_{{\rm c}}$ is the difference in $j_{{\rm c}}$. This
value is almost same as the value obtained in the thermographic experiments
\cite{Daimon:2016fja,Daimon:2017jx}; since in Ref. \cite{Daimon:2017jx}
the sine-wave amplitude $A$ of $\Delta T_{{\rm SPE}}$ is divided
by the rectangular-wave amplitude of $\Delta j_{{\rm c}}$, a correction
factor of $\pi/4$ is necessary, i.e. $\Delta T_{{\rm SPE}}/\Delta j_{{\rm c}}=\pi A/\left(4\Delta j_{{\rm c}}\right)=3.7\times10^{-13}\ {\rm Km^{2}/A}$
in the previous study. We note that, in the above and following measurements,
$t_{{\rm delay}}=50\ {\rm ms}$ is chosen based on the $t_{{\rm delay}}$
dependence of $\Delta T_{{\rm SPE}}$ {[}Fig. \ref{fig:LF-SPE}(c){]},
where $\Delta T_{{\rm SPE}}$ is almost saturated at $t_{{\rm delay}}>10\ {\rm ms}$.
Such finite but small thermal-stabilization time can be explained
by the thermal diffusion from the junction to the TC and rapid thermal
stabilization of the SPE-induced temperature modulation \cite{Daimon:2016fja}. 

\begin{figure*}
\includegraphics{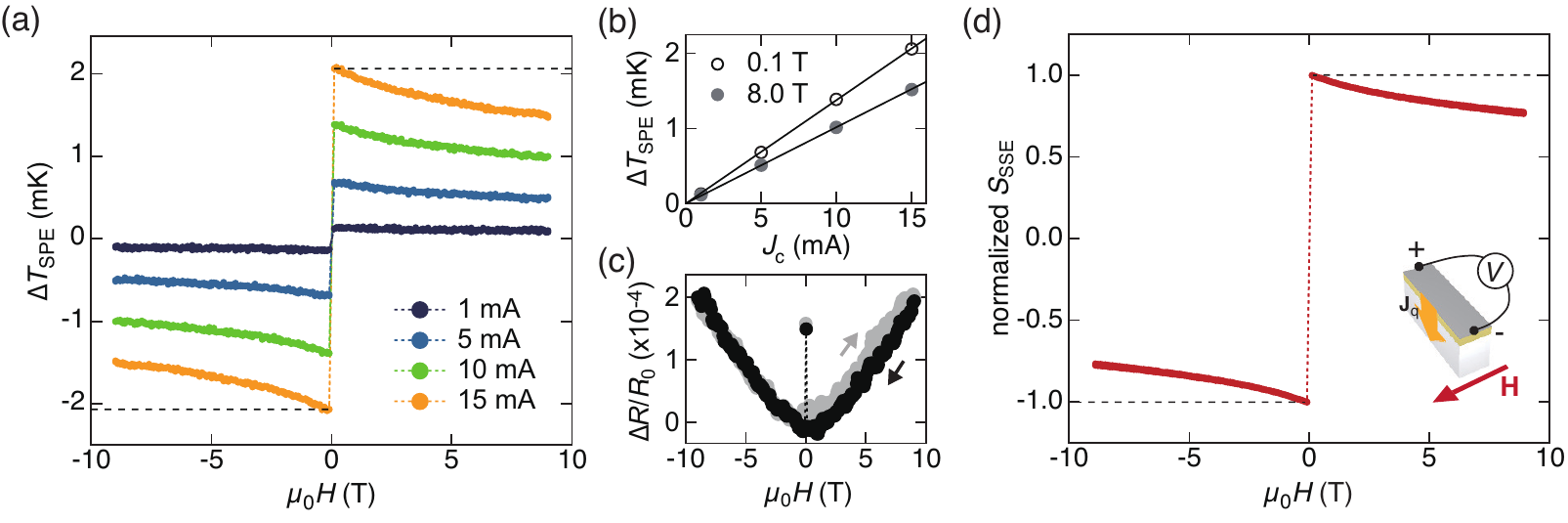}

\caption{(a) Field dependence of $\Delta T_{{\rm SPE}}$ including the high-magnetic-field
range measured at various $\Delta J_{{\rm c}}$ values. (b) $J_{{\rm c}}$
dependence of $\Delta T_{{\rm SPE}}$ at $\mu_{0}\left|H\right|=0.1\ {\rm T}$
and $8.0\ {\rm T}$. The solid lines represent the linear fitting.
(c) Field dependence of the resistance change $\Delta R=R\left(H\right)-R_{0}$,
where $R$ denotes the resistance and $R_{0}$ that at $\mu_{0}H=0.1\ {\rm T}$.
(d) Field dependence of the spin Seebeck voltage $V_{{\rm SSE}}$
normalized at $\mu_{0}H=0.1\ {\rm T}$, where $V_{{\rm SSE}}$ is
obtained by measuring the voltage under heater power of $100\ {\rm mW}$
and subtracting the component symmetric to $H$. \label{fig:SPE_HF}}
\end{figure*}

Next, we measured the field dependence of the SPE at higher fields
up to $\mu_{0}H=9.0\ {\rm T}$. Figure 3(a) shows $\Delta T_{{\rm SPE}}$
as a function of $H$, where $\Delta J_{{\rm c}}$ is changed from
1 to 15 mA. The suppression of the $\Delta T_{{\rm SPE}}$ signal
at higher fields is clearly observed for all the $\Delta J_{{\rm c}}$
values. As shown in Fig. \ref{fig:SPE_HF}(b), the signal shows a
linear variation with $J_{{\rm c}}$ both at $\mu_{0}H=0.1\ {\rm T}$
and 8.0 T, demonstrating a constant suppression rate. Since the resistance
of the sample varies only 1 \% at most {[}Fig. \ref{fig:SPE_HF}(c){]},
the junction temperature keeps constant during the field scan. The
field dependence of the thermal conductivity of YIG is also irrelevant
to the $\Delta T_{{\rm SPE}}$ suppression as it is known to be negligibly
small at room temperature \cite{Boona:2014fh}. Thus we can conclude
that the suppression is attributed to the nature of the SPE. By calculating
the suppression magnitude as 
\[
\delta_{{\rm SPE}}=1-\frac{\Delta T_{{\rm SPE}}\left(\mu_{0}H=8.0\ {\rm T}\right)}{\Delta T_{{\rm SPE}}\left(\mu_{0}H=0.1\ {\rm T}\right)},
\]
we obtained $\delta_{{\rm SPE}}=0.26$.

To compare $\delta_{{\rm SPE}}$ to the field-induced suppression
of the SSE, we performed SSE measurements in a longitudinal configuration
using a Pt/YIG junction system fabricated at the same time as the
SPE sample. The SSE sample has the lateral dimension $2.0\times6.0\ {\rm mm}^{2}$
and the same vertical configuration as the SPE sample except for the
absence of the ${\rm Al_{2}O_{3}}$ layer. The detailed method of
the SSE measurement is available elsewhere \cite{Uchida:2012ew,Anonymous:2017jm,Sola:2017ki}.
Figure \ref{fig:SPE_HF}(d) shows the field dependence of the SSE
thermopower in the Pt/YIG junction. The clear suppression of the SSE
thermopower is observed. Importantly, the high-field response of the
SSE is quite similar to that of the SPE in the Pt/YIG system. The
suppression magnitude of the SSE $\delta_{{\rm SSE}}$, defined in
the same manner as the SPE, is estimated to be $\sim0.22$, consistent
with the previously reported values \cite{Kikkawa:2015bn,Jin:2015ik,Guo:2016go}.

The observed remarkable field-induced suppression of the SPE at room
temperature shows that the SPE is likely dominated by low-energy magnons
because the energy scale of the applied field is less than $10\ {\rm K}$
and thus much lower than the thermal energy of 300 K \cite{Kikkawa:2015bn}.
The origin of the strong contribution of the low-energy magnons in
the SPE can be (i) stronger coupling of the spin torque to the low-energy
(sub-thermal) magnons and (ii) greater propagation length of the low-energy
magnons than those of high-energy (thermal) magnons \cite{Anonymous:2017de}.
While (i) is not well experimentally investigated, the existence of
the ${\rm \mu m}$-range length scale in the SPE \cite{Daimon:2017jx}
and the similarity between $\delta_{{\rm SPE}}$ and $\delta_{{\rm SSE}}$
suggest the dominant contribution from (ii) as in the case of the
SSE \cite{Guo:2016go,Anonymous:2017de}. In fact, recently, it has
been demonstrated that the high magnetic fields reduce the propagation
length of magnons contributing to the SSE \cite{Anonymous:2017de}.
This length-scale scenario can qualitatively explain the suppression
in the SPE. Recalling that a heat current density ($j_{{\rm q}}$)
existing over a distance ($l)$ generates the temperature difference
$\Delta T=\kappa^{-1}j_{{\rm q}}l$ in an isolated system, $\Delta T$
should decrease when $l$ decreases, where $\kappa$ is the thermal
conductivity of the system. In the SPE, $l$ corresponds to the magnon
propagation length \cite{Cornelissen:2016ji}, and a flow of magnons
accompanies a heat current \cite{Boona:2014fh}. Consequently, when
the high magnetic field is applied and the magnons with longer propagation
length are suppressed by the Zeeman gap, the averaged magnon propagation
length decreases and thus results in the reduced $\Delta T$. To further
investigate the microscopic mechanism of the SPE, consideration of
the spectral non-uniformity may be vital both in experiments and theories.

\section{Summary}

In this study, we showed the magnetic field dependence of the spin
Peltier effect (SPE) up to 9.0 T at 300 K in a Pt/YIG junction system.
We established a simple but sensitive detection method of the SPE
using a commonly-available thermocouple wire. The SPE signals were
observed to be suppressed at high magnetic fields, highlighting the
stronger contribution of low-energy magnons in the SPE. The similar
suppression rate of the SPE-induced temperature modulation to that
of the SSE-induced thermopower suggests that the suppression originates
the decrease in the magnon propagation length as in the case of the
SSE. We anticipate that the experimental results and the method reported
here will be useful for systematic investigation of the SPE.
\begin{acknowledgments}
The authors thank T. Kikkawa for the aid in measuring the SSE and
G. E. W. Bauer and Y. Ohnuma for the valuable discussion. This work
was supported by PRESTO \textquotedblleft Phase Interfaces for Highly
Efficient Energy Utilization\textquotedblright{} (Grant No. JPMJPR12C1)
and ERATO \textquotedblleft Spin Quantum Rectification Project\textquotedblright{}
(Grant No. JPMJER1402) from JST, Japan, Grant-in-Aid for Scientific
Research (A) (Grant No. JP15H02012), and Grant-in-Aid for Scientific
Research on Innovative Area \textquotedblleft Nano Spin Conversion
Science\textquotedblright{} (Grant No. JP26103005) from JSPS KAKENHI,
Japan, NEC Corporation, the Noguchi Institute, and E-IMR, Tohoku University.
S.D. was supported by JSPS through a research fellowship for young
scientists (Grant No. JP16J02422). K.O. acknowledges support from
GP-Spin at Tohoku University. 
\end{acknowledgments}

\newpage{}

\appendix

\section{Calibration of Thermo Couple at High Magnetic Fields\label{sec:Calibration-of-Thermo}}

\begin{figure}
\includegraphics{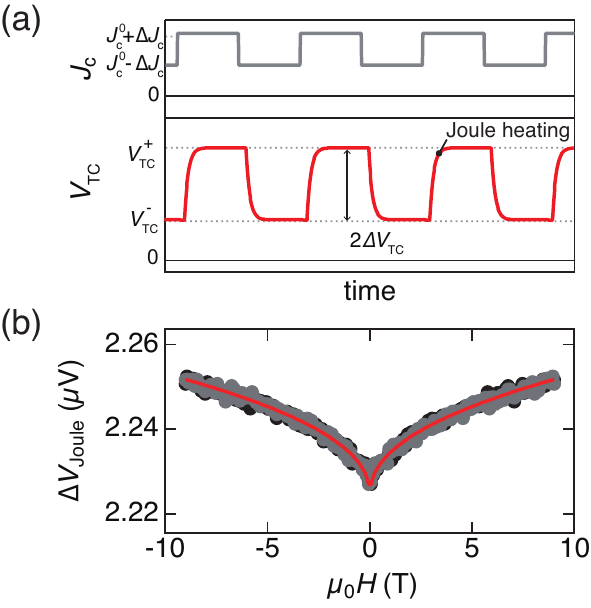}

\caption{(a) Expected responses due to the Joule heating at the finite offset
current $J_{{\rm c}}^{0}\protect\neq0$. (b) Field dependence of $\Delta V_{{\rm Joule}}$
at $\Delta J_{{\rm c}}=0.5\ {\rm mA}$ and $J_{{\rm c}}^{0}=5\ {\rm mA}$.
The solid curve represents the calibration line determined by the
fitting. \label{fig:Schematic-1}}
\end{figure}
To measure the field dependence of $S_{{\rm TC}}$, we used the Joule-heating-induced
signal as a reference. By adding a non-zero offset ($J_{{\rm c}}^{0}$)
to the applied current, we obtained the temperature modulation induced
by the Joule heating, of which the power $P$ changes from $P\left(H\right)=R\left(H\right)\left(J_{{\rm c}}^{0}-\Delta J_{{\rm c}}\right)^{2}$
to $P\left(H\right)=R\left(H\right)\left(J_{{\rm c}}^{0}+\Delta J_{{\rm c}}\right)^{2}$,
where $R$ denotes the resistance of the strip {[}Fig. \ref{fig:Schematic-1}(a){]}.
Figure \ref{fig:Schematic-1}(b) shows the magnetic field dependence
of the component of $\Delta V_{{\rm TC}}$ symmetric to the field
($\Delta V_{{\rm Joule}}=\left[\Delta T_{{\rm TC}}\left(+H\right)+\Delta T_{{\rm TC}}\left(-H\right)\right]/2$).
As the change in $R$, due to the ordinary, spin Hall, and Hanle magnetoresistance
effects \cite{Nakayama:2013gs,Velez:2016bm}, is in the order of 0.02
\% {[}Fig.3(c){]}, its contribution to $P$ can be neglected. Similarly,
the field dependence of the thermal conductivity of YIG is negligibly
small \cite{Boona:2014fh}, ensuring the constant temperature change.
Accordingly, the field dependence of $\Delta V_{{\rm Joule}}$ directly
reflects $S_{{\rm TC}}\left(H\right)$. It increases by a factor of
$\sim1\ \%$ when the field magnitude increases up to 9.0 T. We approximated
the field dependence of $S_{{\rm TC}}\left(H\right)$ as $S_{{\rm TC}}\left(H\right)=61\left(1+4.54\times10^{-3}\left|\mu_{0}H\right|^{0.453}\right)\ {\rm \mu V/K}$
by determining the relative change from the measurement results $\left(\Delta V_{{\rm Joule}}\left(H\right)/\Delta V_{{\rm Joule}}|_{H=0}\right)$
and the absolute value from the reference value. 

\bibliographystyle{apsrev4-1}

\end{document}